\documentstyle[bo99,epsfig]{article}

\title{Spectral Signatures of KiloHertz Quasi-Periodic
Oscillations from Accreting Neutron Stars}

\author{Philip Kaaret}

\affil{Harvard-Smithsonian Center for Astrophysics}

\begin{document}

\maketitle

\begin{abstract}

Correlations discovered between millisecond timing properties
and spectral properties in neutron star x-ray binaries are
described and then interpreted in relation to accretion flows
in the systems.  Use of joint timing and spectral
observations to test for the existence of the marginally
stable orbit, a key prediction of strong field general
relativity, is described and observations of the neutron star
x-ray binary 4U~1820-303 which suggest that the signature of
the marginally stable orbit has been detected are presented.


\end{abstract}

\section{Introduction}

The orbital periods associated with the innermost orbits
around solar mass compact objects are in the millisecond
range.  The millisecond quasi-periodic oscillations (QPOs)
discovered (Strohmayer et al.\ 1996; van der Klis et al.\
1996) with the {\it Rossi X-Ray Timing Explorer} (RXTE;
Bradt, Rothschild, \& Swank 1993) in the x-ray emission from
accreting neutron stars are most likely associated with
orbital motion near the neutron star, and, thus, potentially
offer a new probe of accretion flows and strong gravitational
fields near neutron stars in x-ray binaries (for a brief
review, see Kaaret \& Ford 1997).  The key to exploiting the
kHz QPOs as probes of strong gravity is to understand the
QPOs at a sufficient level so that those aspects of the QPOs
which depend on the properties of the gravitational field can
be separated from the aspects which depend on details of the
accretion physics.  It may not be necessary to construct a
full theory of the QPO generation mechanism, just as a binary
pulsars are excellent test beds for the study of relativity
even in the absence of an adequate theory for the generation
of the radio pulses.

Here, I present some initial steps of attempts to understand
the key quantitative features of the QPOs, such as the
centroid frequency, and their relation to the properties of
the accretion flow.  I consider several key questions:

\begin{enumerate}

\item Is the x-ray spectral state of a source correlated with
QPO frequency?  

\item Is the QPO frequency related to the geometry of an
accretion disk?

\item Is the QPO frequency determined by the mass accretion
rate?

\end{enumerate}

In the following, I present and then interpret observational
data to attempt to answer these three questions.    

If an adequate understanding of the QPOs can be obtained, we
can then extend our study to address the properties of the
gravitational field surrounding the neutron star.  One
important question is whether there exists an innermost
radius, the marginally stable orbit also referred to as the
innermost stable circular orbit, inside of which there exist
no stable circular orbits.  The lack of stability of circular
orbits sufficiently close to a compact object is a key
prediction of strong field general relativity (e.g. Misner,
Thorne, \& Wheeler 1970).  The relevance of the existence of
the marginally stable orbit to to the physics of accretion
around black holes has long been understood (e.g. Shakura \&
Sunyaev 1973).  More recently, it was pointed out by Kluzniak
and Wagoner (1985) that, for most of the preferred equations
of state of nuclear matter, the marginally stable orbit lies
outside the neutron star radius and thus should be
dynamically important for the accretion flow on to neutron
stars.  The kHz QPOs appear to provide an observational means
to test these predictions.

\section{Correlation of QPO frequency with spectral
state}

We wish to use the kHz QPOs as kinematic probes and, thus,
must understand the relation of the QPO parameters,
particularly the centroid frequency, to the geometry of the
accretion flow.  We begin by searching for correlations in
the time variations of QPO parameters and x-ray spectral
properties.

\begin{figure} \centerline{\psfig{file=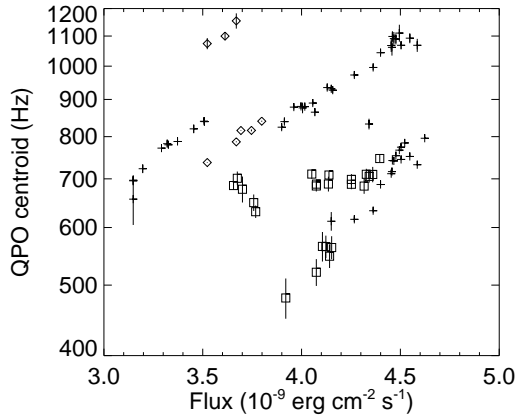,
width=8cm}} \caption[]{QPO centroid frequency versus x-ray
flux in the 2.5--25~keV band for 4U~1728-34 (GX 354-0). 
Included are data from observations spanning three years. 
The different symbols indicate data from different years.}
\end{figure}

The most obvious spectral parameter with which to correlate
the QPO frequency is the total X-ray flux.  In a neutron star
system, accreted matter must eventually come to rest on the
neutron star surface, and, therefore, the total luminosity,
$L$, of a neutron star system must be directly proportional
to the total mass accretion rate onto the stellar surface
$\dot{M}$ via $L = G M \dot{M} / R$, where $G$ is the
gravitational constant, and $M$ and $R$ are the mass and
radius of the neutron star.  If the emission is isotropic,
or, more generally, if the beaming pattern of the emission
does not change with time, then the bolometric flux from a
source must also be proportional to $\dot{M}$.  In Fig.~1, we
show the relation between QPO frequency and X-ray flux in the
2.5--25 keV band for the neutron star binary 4U~1728-34 (GX
354-0).  The figure includes data from observations spanning
three years.  It is apparent from the figure that, while the
QPO frequency and X-ray flux may be correlated for subsets of
the data, the overall behavior shows no correlation.  The
lack of correlation between QPO frequency and X-ray flux may
be due to one or more of several different factors: time
variable beaming of the X-rays, the contribution of
significant flux outside the measurement band, an outflow
from the system, or a fundamental lack of physical
correlation between QPO  frequency and total mass accretion
rate.

Spectral shape has long been employed as an indicator of mass
accretion rate in neutron star systems.  For sources with
x-ray spectra which can be adequately described with a simple
power law model (often used for black hole candidates and low
luminosity neutron star binaries), a natural indicator of
spectral state is the power law photon index.  Kaaret et al.\
(1998) showed that the QPO frequency is well correlated with
photon index, but not with x-ray flux, in the neutron star
x-ray binaries 4U~0614+091 and 4U~1608-52.  This suggests
that the spectral state and QPO frequency are related to a
common physical parameter. Kaaret et al.\ (1998) suggested
that common parameter is the mass accretion rate through the
accretion disk.  

\begin{figure} \centerline{\psfig{file=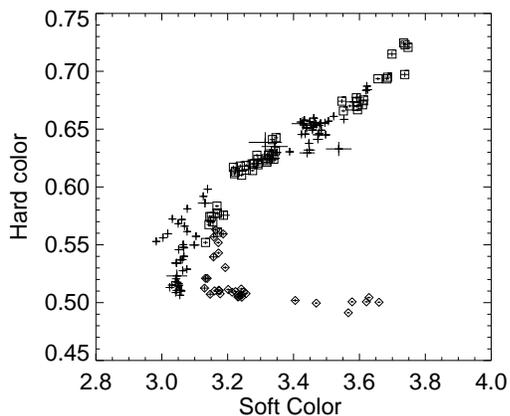,
width=8cm}} \caption[]{Color-color diagram for 4U~1728-34 (GX
354-0).  The soft color is defined as the ratio of counts in
the 3.5--6.4~keV band to counts in the 2.0--3.5~keV band and
the hard color as 9.7--16.0~keV to 6.4--9.7~keV.  The plot
symbols are the same as in Fig.~1.} \end{figure}

\begin{figure} \centerline{\psfig{file=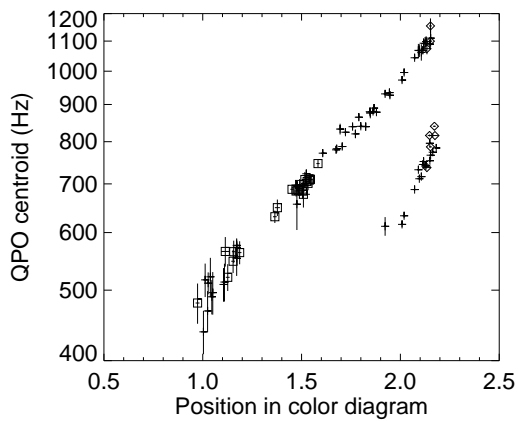,
width=8cm}} \caption[]{QPO centroid frequency versus position
in color-color diagram for 4U~1728-34.  Larger values of the
position in the color-color diagram indicate higher inferred
mass accretion rates.  The upper and lower branches
correspond to the upper and lower frequency kHz QPOs which
often appear simultaneously.  The plot symbols are the same
as in Fig.~1.} \end{figure}

A more general indicator of spectral state is an x-ray color,
defined as the ratio of count rates in two different energy
bands.  X-ray colors can be applied to any source and do not
rely on an assumed spectral model.  However, they are
strongly instrument dependent, making it difficult to compare
colors obtained from different instruments, and they have no
direct interpretation in terms of physical parameters of the
source.  X-ray colors have been used as spectral state
indicators both individually and in pairs forming a
``color-color'' diagram.  When their time varying spectral
state is plotted on a color-color diagram, each individual
neutron star x-ray binary tends to follow a well-defined
track, see Fig.~2 where we present a color-color diagram for
4U~1728-34.  This collapse of a potentially two dimensional
pattern to a one-dimensional track suggests that a single
parameter determines the spectral state.  This parameter has
usually been interpreted as the total mass accretion rate in
the system (Hasinger \& van der Klis 1989).  However, the
lack of correlation between total flux and spectral state may
indicate that the parameter is not the total mass accretion
rate.

Position in the color-color diagram can be parameterized by
position along a fiducial track drawn through the diagram
(Hasinger et al.\ 1990; Hertz et al.\ 1992).  In Fig.~3, we
show the QPO frequency versus position in the color-color
diagram (Fig.~2) for the same observations presented in
Fig.~1.  The correlation between QPO frequency and position
in the color-color diagram appears robust across several
years of observations.  A robust correlation is found between
QPO frequency and x-ray colors, either position in a
color-color diagram or simply a hard x-ray color, for most
neutron star x-ray binaries (e.g. Mendez 1999).

\begin{figure} \centerline{\psfig{file=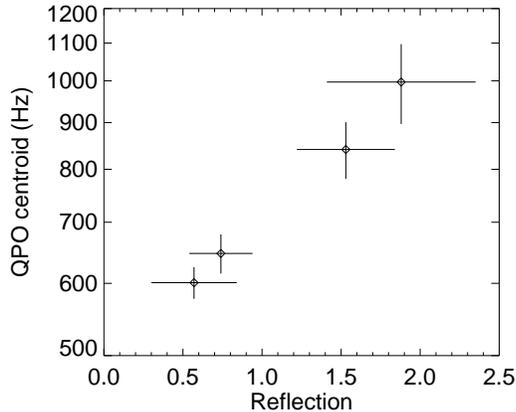,
width=8cm}} \caption[]{QPO frequency versus magnitude of
reflection for 4U~0614+091. The inclination angle for the
reflection component was fixed to $0^{\circ}$.} \end{figure}

\section{QPO frequency and reflection}

The robust correlation between QPO centroid frequency and
various indicators of spectral state, photon index or x-ray
colors, suggest that a single physical parameter determines
both spectral state and QPO frequency.  To determine the
physical nature of this parameter, we performed a detailed
analysis of high quality x-ray spectra of 4U~0614+091
obtained with BeppoSAX (Piraino et al.\ 1999).  When fit with
a simple power-law model, the spectra showed strong and
systematic residuals near 10--30~keV with a shape
characteristic of that expected for a reflection component in
the spectrum.  Addition of reflection to the spectral model
greatly improved the quality of the fits and gave residuals
with no systematic variations.  Thus, there is strong
evidence for a reflection component in the spectra of
4U~0614+091.  We found that the magnitude of reflection is
well correlated with photon index, confirming a relation
suggested by Zdziarski et al.\ (1999).

The reflection component found in the spectra requires the
presence of cool matter located close to the primary x-ray
source, presumably the neutron star, and subtending a large
solid angle as viewed by the primary source.  The most
natural interpretation is that the reflection occurs in an
accretion disk surrounding the neutron star.  In this case,
the magnitude of reflection should be related to the
properties of the disk.  In particular, if the disk has a
variable inner radius, then higher magnitudes of reflection
will result for smaller disk inner radii.

Fig.~4 shows the QPO centroid frequency plotted versus
magnitude of reflection for 4U~0614+091.  The QPO parameters
are from RXTE observations while the magnitude of reflection
is from BeppoSAX observations.  For the two points with lower
frequency, we had simultaneous RXTE and BeppoSAX
observations. For the upper two points, we inferred the QPO
frequency based on the QPO frequency versus photon index
relation from Kaaret et al.\ (1998) and the photon index
measured with BeppoSAX and allowing for a systematic offset
in photon indices between BeppoSAX and RXTE measured using
simultaneous observations.  The large error bars for these
two points are due to the uncertainty in this extrapolation.
The QPO frequency appears correlated with the magnitude of
reflection.  The correlation is consistent with that expected
if the QPO frequency is determined by the orbital frequency
at the inner edge of the disk and the variation in reflection
is due to changes in the inner disk radius.

\section{kHz QPOs and accretion geometry}

I suggest the following physical picture to explain the
correlations of QPO frequency with spectral state and
magnitude of reflection presented in the previous sections. 
Mass accretion can occur in neutron star x-ray binaries both
through the accretion disk and radially (e.g. Ghosh \& Lamb
1978).  The total mass accretion rate, disk plus radial,
determines the total luminosity of the system.  The mass
accretion rate through the disk determines the radius of the
inner edge of the disk and, via the dependence of the overall
spectrum on the soft photon flux emitted from the disk, the
spectral state of the system.  The QPO frequency is
determined by the radius of the inner edge of the disk. The
data presented above are fully consistent with this picture.

To answer the questions posed earlier:

\begin{enumerate}

\item The spectral state of a source is correlated with QPO
frequency as demonstrated by the robust correlations found
between QPO frequency and photon index or x-ray color.

\item The QPO frequency is related to the geometry of the
disk as demonstrated by the correlation between QPO frequency
and magnitude of reflection in the x-ray spectrum.

\item The QPO frequency is determined by the mass accretion
rate, but the important parameter is the mass accretion rate
through the disk and not the total mass accretion rate.

\end{enumerate}

\begin{figure} \centerline{\psfig{file=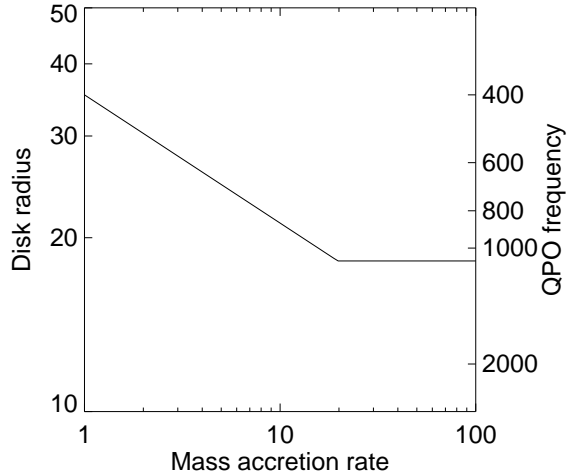,
width=8cm}} \caption[]{Disk radius versus mass accretion rate
through the disk showing the effect of the marginally stable
orbit.  The disk radius, QPO frequency, and mass accretion
rate are all in arbitrary units.} \end{figure}

\section{kHz QPOs and the marginally stable orbit}

The existence of the marginally stable orbit will have a
strong effect on the configuration of the accretion disk near
the neutron star as the lack of stable orbits inwards of the
marginally stable orbit implies that a stable disk can not
exist in that region.  The inner radii of disks around
neutron stars appear to be variable.  The truncation of the
disk may be caused by the neutron star magnetic field, by
radiation forces acting on the disk, or by a disk
instability.  In general, the inner disk radius decreases
with increasing mass accretion rate through the disk.  The
marginally stable orbit will modify this behavior by limiting
the minimum possible inner disk radius, see figure~5.  Thus,
saturation at a minimum disk radius for large mass accretion
rates is a signature of the marginally stable orbit (Miller,
Lamb, \& Psaltis 1998).  As described in Kaaret, Ford, \&
Chen (1997), when the disk reaches the marginally stable
orbit, the inner radius will not approach a single value, but
instead wander over some range due to the properties of the
transonic flow near the marginally stable orbit.

If one is willing to accept the assertions made in the
previous section that the QPO frequency is determined by the
radius of the inner edge of the accretion disk and that the
x-ray spectral state is an indicator of mass accretion rate
through the disk, then the relation of QPO frequency versus
spectral state can be used to probe the relation of inner
disk radius versus mass accretion rate through the disk. 
Thus, the shape of the QPO versus spectral state diagram,
i.e. whether or not the QPO frequency saturates at high mass
accretion rates, provides a test for the presence of the
marginally stable orbit.

\begin{figure} \centerline{\psfig{file=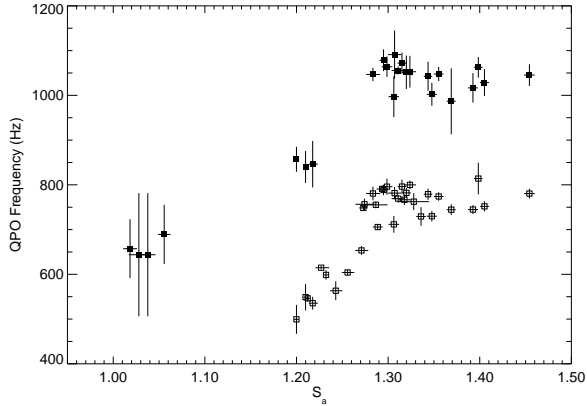,
width=8cm}} \caption[]{QPO centroid frequency versus position
in color-color diagram for 4U~1820-30 from Bloser et al.\
(1999).  Larger values of the position in the color-color
diagram indicate higher inferred mass accretion rates. 
Filled squares indicate the upper kHz QPO while open squares
indicate the lower kHz QPO.} \end{figure}

The QPO frequency versus spectral state diagram for
4U~1728--34, Fig.~3, shows no evidence for a saturation of
QPO frequency at high mass accretion rates.  This is also
true for almost all neutron star binary for which kHz QPOs
have been detected.  There is only one source which does show
a clear saturation of QPO frequency at high inferred mass
accretion rates: 4U~1820-30.  Zhang et al.\ (1998)
demonstrated a saturation of QPO frequency versus x-ray count
rate for 4U~1820-30.  However, as QPO frequency is poorly
correlated with count rate in most sources (the relation
tends to look similar to the QPO frequency versus flux
relation in Fig.~1), this approach came under significant
criticism. Kaaret et al.\ (1999) showed that the QPO
frequency saturates when plotted versus a spectral state
indicator derived from a hard x-ray color.  As QPO frequency
is generally well correlated with hard x-ray color (e.g.
Mendez 1999), this result is compelling evidence  that the
QPO frequency saturates at high disk mass accretion rates. 
Recently, Bloser et al.\ (1999) performed a similar analysis
using position in a color-color diagram as a spectral state
indicator, see Fig.~6, and again find a saturation of QPO
frequency.

\begin{figure} \centerline{\psfig{file=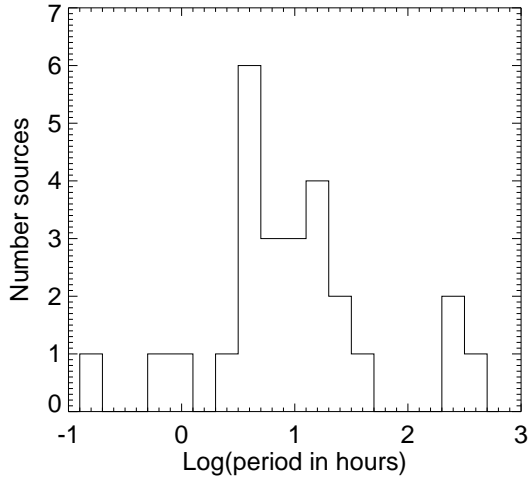,
width=8cm}} \caption[]{Period distribution of low-mass x-ray
binaries from the catalog of van Paradijs (1995). 4U~1820-30
has the shortest known orbital period at 11~minutes.}
\end{figure}

The fact that only 4U~1820-30 exhibits a saturation of the
QPO frequency versus inferred disk mass accretion rate, and
thus appears to be the only source for which the disk reaches
the marginally stable orbit, merits some consideration.
Interestingly, 4U~1820-30 also holds a unique position as the
stellar binary system with the shortest known orbital period,
see Fig.~7, and is thought to be a double degenerate binary
which evolved through a common envelope phase (Rappaport et
al.\ 1987).  These unique features of 4U~1820-30 suggest that
the neutron star may have accreted more matter and thus is
more massive than other neutron stars in x-ray binaries.

For the marginally stable orbit to be dynamically important,
it must lie outside the neutron star surface.  If there is a
boundary layer on the neutron star surface formed by the
accretion flow, then to be observed via kHz QPOs, the
marginally stable orbit must lie outside the top surface of
the boundary layer.  Inogamov \& Sunyaev (1999) have recently
calculated the properties of such boundary layers and found
typical equatorial thicknesses near 1~km.  This is comparable
to the separation between the marginally stable orbit and
neutron star surface for a $1.4 \, M_{\odot}$ neutron star,
rotating at 250--400~Hz, for many of the currently favored
equations of state for nuclear matter.  The separation
increases as the neutron star mass increases since the
surface generally moves inward and the marginally stable
orbit moves outward.  Thus, the signature of the marginally
stable orbit is most likely to be observed from the most
massive neutron stars.

\section{Conclusions}

Millisecond x-ray timing is a promising probe of accretion
flows in x-ray binaries.  Robust correlations exist between
millisecond timing properties and spectral properties. Study
of these correlation may help improve our understanding of
the physics and geometry of accretion flows in neutron star
x-ray binaries and, perhaps, also of strong field gravity.

\begin{acknowledgements}

I thank Peter Bloser for use of Fig.~6 before publication.

\end{acknowledgements}


\begin{references}

\ref Bloser, P.F., Grindlay, J.E., Kaaret, P., Zhang, W.,
Smale, A.P., \& Barret, D.\ 2000, ApJ to appear,
astro-ph/0005496

\ref Bradt, H.V., Rothschild, R.E., \& Swank, J.H. 1993, AAS,
97, 355

\ref Ghosh, P., \& Lamb, F. K. 1979, ApJ, 234, 296 

\ref Hasinger, G. \& van der Klis, M. 1989, A$\&$A, 225, 79

\ref Hasinger, G., van der Klis, M., Ebisawa, K., Dotani, T.,
\& Mitsuda, K. 1990, A$\&$A, 235, 131

\ref Hertz, P., Vaughan, B., Wood, K.S., Norris, J.P.,
Mitsuda, K., Michelson, P.F., \& Dotani, T. 1992, ApJ,  396,
201

\ref Inogamov, N.A. \& Sunyaev, R.A. 1999, Astron. Lett. 25,
269

\ref Kaaret, P., Ford, E. C. \& Chen, K. 1997, ApJ, 480, L27

\ref Kaaret, P. \& Ford, E.C. 1997, Science, 276, 1386

\ref Kaaret, P., Yu, W., Ford, E.C., \& Zhang, S.N. 1998,
ApJ, 497, L93

\ref Kaaret, P. et al.\ 1999, ApJ, 520, L37
 
\ref Kluzniak, W. \& Wagoner, R.V.  1985, ApJ, 297, 548

\ref Mendez, M. 1999, Proceedings of the 19th Texas Symposium
in Paris, astro-ph/9903469

\ref Miller, M. C., Lamb, F. K., \& Psaltis, D. 1998,
ApJ, 508, 791

\ref Misner, C.W., Thorne, K.S., \& Wheeler, J.A. 1970,
Gravitation (San Francisco: Freeman)

\ref Piraino, S., Santangelo, A., Ford, E.C., \& Kaaret, P.
1999, A$\&$A, 349, L77

\ref Rappaport, S., Ma, C.P., Joss, P.C., Nelson, L.A. 1987,
ApJ, 322, 842

\ref Shakura, N.I. \& Sunyaev, R.A. 1973, A$\&$A, 24, 337

\ref Strohmayer, T.E. et al.\ 1996, ApJ, 469, L9

\ref van der Klis, M. et al.\ 1996, ApJ, 469, L1

\ref van Paradijs, J.\ 1995, in X-Ray Binaries, ed. W.H.G.\
Lewin, J.\ van Paradijs, \& P.J.\ van den Heuvel, (Cambridge:
Cambridge University Press), 536

\ref Zdziarski, A.A., Lubiski, P., Smith, D.A. 1999, MNRAS,
303, L11

\ref Zhang, W., Smale, A.P., Strohmayer, T.E., \& Swank, J.H.
1998, ApJ, 500, L171

\end{references}
\end{document}